\documentclass[12pt, authoryear]{elsarticle}
\usepackage{amssymb}
\usepackage{amsmath}
\usepackage{amsfonts}
\usepackage{graphicx}
\usepackage[utf8]{inputenc}
\usepackage[english]{babel}
\usepackage[table,xcdraw]{xcolor}
\usepackage{subcaption}
\usepackage{booktabs}
\usepackage[markers,nolists]{endfloat}

\newcommand{\on}{\operatorname}
\newcommand{\rv}{\tilde}
\newcommand{\R}{\mathbb{R}}
\newcommand{\N}{\mathcal{N}}
\newcommand{\rpm}{\raisebox{.2ex}{$\scriptstyle\pm$}}

\begin{document}
\begin{frontmatter}
\title{Groupwise Structural Parcellation of the Cortex: A~Sound Approach Based on Logistic Models}
\author[nice]{Guillermo~Gallardo}
\author[harvard]{William~Wells~III}
\author[nice]{Rachid~Deriche}
\author[nice]{Demian~Wassermann}
\address[nice]{Universit\'e C\^ote d'Azur, Inria, France}
\address[harvard]{Harvard Medical School, Boston, Massachusetts, USA}
\begin{abstract}
Current theories hold that brain function is highly related to long-range
physical connections through axonal bundles, namely \textit{extrinsic connectivity}.
However, obtaining a groupwise cortical parcellation based on extrinsic connectivity
remains challenging. Current parcellation methods are computationally expensive;
need tuning of several parameters or rely on ad-hoc constraints. Furthermore,
none of these methods present a model for the cortical extrinsic connectivity
of the cortex.
To tackle these problems, we propose a parsimonious model for the 
extrinsic connectivity and an efficient parceling technique based on
clustering of tractograms. Our technique allows the creation of single subject
and groupwise parcellations of the whole cortex. The parcellations obtained with
our technique are in agreement with structural and functional parcellations
in the literature. In particular, the motor and sensory cortex are subdivided
in agreement with the human homunculus of Penfield. We illustrate this by comparing
our resulting parcels with the motor strip mapping included in the Human
Connectome Project data.
\end{abstract}
\end{frontmatter}
\section{Introduction}
The human brain is arranged in areas based on criteria such as cytoarchitecture,
functional specialization or axonal connectivity~\citep{Brodmann1909, Thirion2014,
ThiebautdeSchotten2016}. Parceling the cortex into such areas and 
characterizing their interaction is key to understanding how the brain works.
Nowadays it's accepted that axonal connectivity plays a fundamental role in the
interaction between brain regions~\citep{Schmahmann2006}. Moreover, current theories
hold that long-range physical connections trough axonal bundles,
namely \textit{extrinsic connectivity}, are strongly related to brain function, for example,
this has been shown in macaques~\citep{Passingham2002}. Therefore, understanding
how the cortex is arranged based on its extrinsic connectivity can
provide key information in unraveling the internal organization of the brain.

Diffusion MRI (dMRI) enables the in vivo exploration of extrinsic connectivity 
and other aspects of white matter anatomy on the brain. However, in using diffusion
MRI to infer long-distance connectivity, several challenges arise. A primary issue
is the spatial resolution of diffusion imaging: it is several
orders of magnitude coarser than axonal diameters (millimeters vs.
micrometers)~\citep{VanEssen2014}, making hard to infer some brain pathways.
In addition, there is as yet no quantitative
measure of the strength of connections from diffusion~\citep{Jbabdi2013}.
Given these general limitations, obtaining a cortical parcellation based on
extrinsic connectivity remains challenging~\citep{VanEssen2014, Jbabdi2013}.
Moreover, most current parceling
techniques compute either single-subject or groupwise parcellations.
Single-subject techniques work by refining other
parcellations~\citep{Clarkson2010}, which introduces a bias in the resulting
parcellation; parceling only part of the 
cortex~\citep{Lefranc2016, Roca2009, ThiebautdeSchotten2014, ThiebautdeSchotten2016}
or using ad-hoc metrics to compare extrinsic connectivity~\citep{Moreno-Dominguez2014}.
Meanwhile, existing groupwise methods rely on average connectivity
profiles~\citep{Clarkson2010, Roca2010}, which prevents obtaining single
subject parcellations; seek a matching across subjects after independent
parcellations~\citep{Moreno-Dominguez2014}, relying on possible noisy results,
or need fine tuning of parameters, as the expected number of clusters to
find~\citep{Paristot2015}.

\begin{figure}
    \includegraphics[width=\textwidth]{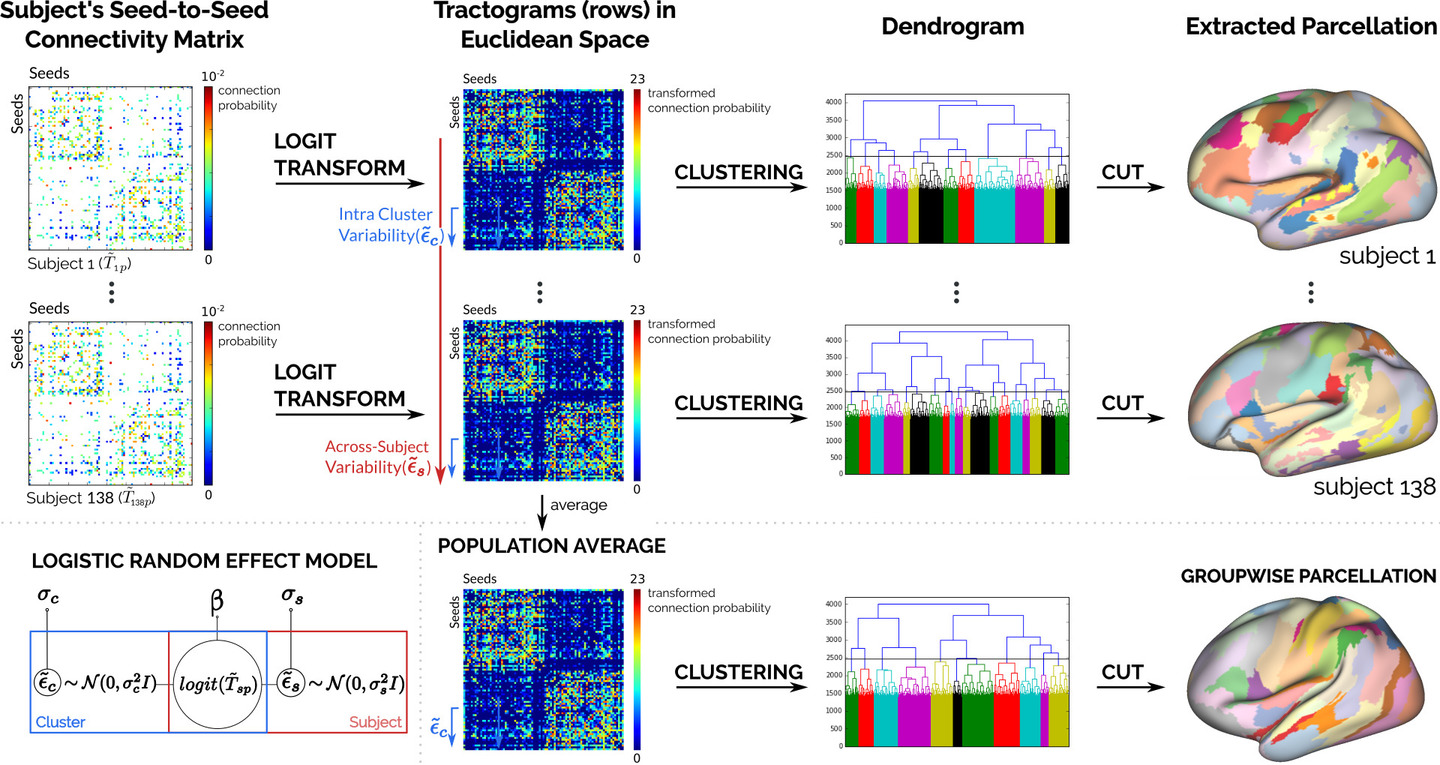}
    \caption{Lower left corner: graphical model of the linear relationship
             between the tractogram of a subject $s$ for a seed $p$ ($\tilde T_{sp}$); and the intra-cluster
             ($\tilde \epsilon_c$) and across-subject ($\tilde \epsilon_s$) variability of the seed's patch. We transform the tractograms 
             into a Euclidean space while explicitly accounting for the variability. This allows us to use well known clustering techniques and 
             compress different levels of granularities for a same parcellation
             in a dendrogram.}
    \label{fig:summary}
\end{figure}

In this work, we present a parsimonious model for the cortical
connectivity alongside an efficient parceling technique based on it. We summarize
both contributions in Fig.~\ref{fig:summary}. Our model assumes that the cortex is
divided in patches of homogeneous extrinsic connectivity. That is, nearby
neurons in the cortex share approximately the same long-range physical
connections, we call this the \textit{local coherence criterion}. Our assumption
is based on histological results in the macaque brain~\citep{Schmahmann2006}.
Inspired by statistical models for clustered data~\citep{Pendergast1996}, our model
accounts for the variability in the axonal connections of neurons within a patch
and for variability in patch boundaries across subjects. Our parceling technique
allows us to create single subject and groupwise parcellations of the whole
cortex in agreement with extant parcellations.

We validate our technique by taking advantage of data available from the Human
Connectome Project (HCP). Using our technique, we compute single subject and a
groupwise parcellations. In this work we will focus on the groupwise case. For
results of our method on the single-subject case
please refer to \citet{Gallardo2016}. Here, we first assess the consistency of our groupwise
parceling technique by comparing the groupwise parcellations of three disjoint
groups of 46 subjects from the HCP. We also show that our technique computes a
similar parcellation to the one obtained by \citet{ThiebautdeSchotten2016} when
parceling only the frontal cortex. Later, to test the functional specialization
of our frontal lobe parcels, we use a data-base of meta-analysis of fMRI studies
\citep{Yarkoni2011}, as in \citet{ThiebautdeSchotten2016}. After, we show
that our groupwise parcels subdivide some well-known anatomical structures
by comparing our results against Desikan's atlas~\citep{Desikan2006}. Also,
we show the functional specialization of some of our parcels by comparing against
results from \citet{Glasser2013}. Finally, we compare our groupwise parcellation
of 138 subjects against the multi-modal parcellation of \citet{Glasser2016}. We
show that, while the parcellations boundaries differ, our parcels show similar
or better functional specialization, specially for motor related tasks.

This work is organized as follows: In the Methods section we present our
model for cortical connectivity and frame tractography within our model. Also,
we present both our single-subject and groupwise case methodologies to parcellate
the cortex. In the Experiments and Results section we present our results on HCP
data. We then discuss our results and position ourselves with respect to the state
of the art in the Discussion section. Finally, in the last section we provide our
conclusions.
\section{Methods}
\subsection{Cortical Connectivity Model and Tractography}
\label{sec:cortical_model}
Our model assumes that the cortex is divided in clusters of homogeneous extrinsic
connectivity. That is, nearby neurons in the cortex share approximately the same
long-ranged physical connections, we call this the \textit{local coherence criterion}.
Our assumption is based on histological results in the macaque brain~\citep{Schmahmann2006}.
As in clustered data models in statistics~\citep{Pendergast1996}, we allow
intra-cluster and across-subject variability in the connectivity. We formalize this concept as:
\begin{equation}
    \label{eq:notation}
    K = \bigcup_{i=1}^k K_i , \forall_{1 \leq i,j \leq k}, i \neq j \rightarrow K_i \cap K_j = \emptyset \land \on{conn}(K_i) \neq \on{conn}(K_j)
\end{equation}
where the set of points on the cortex $K$ is the disjoint union
of each cluster $K_i$ and $\on{conn}(\cdot)$ is the extrinsic connectivity
fingerprint of a cluster. We will make the notion of variability explicit in eq. 
\ref{eq:tractogram_rv}. In this work, the connectivity fingerprint of a seed-point
in the brain is a binary vector denoting to which other seed-points it is
connected through axonal bundles. That is, the physical connections of a point
$p \in K_i$ in the brain are represented by its connectivity fingerprint
$\on{conn}(p) = \on{conn}(K_i)$.

Currently, the most common tool for estimating the extrinsic connectivity
fingerprint of a point in vivo is probabilistic tractography~\citep{Jbabdi2013}.
Given a seed-point in the brain, probabilistic tractography creates a
\textit{tractogram}: an image where each voxel is valued with its probability
of being connected to the seed through axonal bundles. One way of calculating
these probabilities is with a Monte Carlo procedure, simulating the random walk
of water particles through the white matter~\citep{Behrens2003a}. Each one of
these paths is known as a streamline. If we model these streamlines as Bernoulli
trials, where we get a value for the connection from our seed with other points
(1 if they connected by the streamline, 0 if not)~\citep{Behrens2003a}, then, we
can model the tractogram of the subject $s$ in the seed-point $p$ as:
\begin{equation}
    \label{eq:tractogram}
    T_{sp} = 
      [P(\rv C_{spi}=1)]_{1 \leq i \leq n} =
      [\theta_{spi}]_{1 \leq i \leq n}, ~~ \rv C_{spi} \sim \on{Bernoulli}(\theta_{spi})
    \enspace,
\end{equation}
where $\rv C_{spi}$ is a Bernoulli random variable\footnote{For the sake of
clarity we denote all random variables with a tilde, e.g. $\rv C$.} 
representing ``the point $p$ of the subject $s$ is connected to the voxel $i$".
Each Bernoulli's parameter ($\theta_{spi}$) represents the probability of being
connected, and is estimated as the proportion of success in the Bernoulli
trials of each seed.

To formulate the tractogram in accordance to our hypothesis of cortical
connectivity, we model it as a vector of random variables. In our
model, each element in a tractogram comes from a random variable depending on
the point's cluster along with its intra-cluster and across-subject variability:
\begin{equation}
    \label{eq:tractogram_rv}
        p \in K_c \rightarrow
        \rv T_{sp} = 
        [P(\rv C_{spi}=1 | \on{conn}(K_c), ~\rv \epsilon_{ci}, ~\rv \epsilon_{si})]_{1 \leq i \leq n}
        \enspace ,
\end{equation}
in this case, the point $p$ belongs to the cluster $c$;  $\rv \epsilon_{ci}$ 
represents the intra-cluster variability and $\rv \epsilon_{si}$ represents the
across-subject variability for the connectivity to voxel $i$ in the cluster $c$. 

Since each $\rv C_{spi}$ follows a Bernoulli distribution (Eq. \ref{eq:tractogram})
it's difficult to find an explicit formulation for 
$P(\rv C_{spi} = 1 | \on{conn}(K_c),~\rv \epsilon_{ci}, ~\rv \epsilon_{si})$ 
accounting for the variabilities. For this, we use the generalized linear
model (GLM) theory. In this theory, the data is assumed to follow a linear form
after being transformed with an appropriate link function~\citep{McCullagh1989}.
Using the following notation abuse:
\begin{equation}
    \label{eq:not_abuse}
    \on{logit}(\rv T_{sp}) \triangleq  [\on{logit}(P(\rv C_{spi}=1 | \on{conn}(K_c), ~\rv \epsilon_{ci}, ~\rv \epsilon_{si})]_{1  \leq i \leq n},
\end{equation}
\noindent
we derive from GLM a logistic random-effects model~\citep{Pendergast1996} for
each point $p$:
\begin{equation}
    \label{eq:ran_eff_model}
    \on{logit}(\rv T_{sp}) = \beta_{c} + \rv \epsilon_{c} + \rv \epsilon_{s} \in \R^n,
    \quad
    ~ \rv \epsilon_{c} \sim \N(\vec 0, \sigma_c^2 Id),
    ~ \rv \epsilon_{s} \sim \N(\vec 0, \sigma_s^2 Id),
\end{equation}
where $\epsilon_{c}$ and $\epsilon_{s}$ represent the intra-cluster and 
across-subject variability respectively. According to GLM theory 
$\beta_c \in \R^n$ is the extrinsic connectivity fingerprint of cluster $K_c$
transformed: 
\begin{equation}
    \on{logit}^{-1}(\beta_c) = E(\rv T_{sp}) = \on{conn}(K_c) \enspace.
\end{equation}

The choice of logit as link function is based on the work of \citet{Pohl2007}.
In their work, \citet{Pohl2007} show that logit function's codomain is a
Euclidean space, which allows us to transform and manipulate the tractograms 
in a well-known space.
\subsection{Single Subject and Groupwise parceling Methodologies}
\label{sec:parceling_methodologies}
In the previous section, we hypothesized that the cortex is divided in
clusters with homogeneous extrinsic connectivity, alongside intra-cluster and
across-subject variability. In using
the previous hypothesis, it is important to remark that we don't have a priori
knowledge of the cluster's location or their variability. But, thanks to the
proposed logistic random effects model, we formulated the problem of finding
these clusters as a well-known clustering problem. This is because, after
transforming the tractograms with the logit function as in eq.~\ref{eq:not_abuse}
they will be in a Euclidean space~\citep{Pohl2007}. Even more, eq.~\ref{eq:ran_eff_model}
states that the transformed tractograms come from a mixture of Gaussian 
distributions, e.g. it is a Gaussian mixture model.

To solve the Gaussian mixture model and find the clusters, we use a modified
Agglomerative Hierarchical Clustering (AHC) algorithm. This was inspired by the
method of \citet{Moreno-Dominguez2014}. To enforce the local coherence
criterion we also modify the algorithm to accept one parameter: the minimum size
of the resulting clusters. Clusters smaller than this size are merged with
neighbors, i.e. physically close clusters in the cortex. As we are working in
a Euclidean space, we use Ward's Hierarchical Clustering
method~\citep{WardJr.1963}. 
This method creates clusters with minimum within-cluster variance.
The method's result is a dendrogram: a structure that comprises different levels of
granularity for the same parcellation. This allows us to explore different
parcellation granularities by choosing cutting criteria, without the need of
recomputing each time.

The main advantage of the model we proposed in this work is that it
allows us to create a groupwise parcellation using linear operations. Assuming
direct seed correspondence across subjects, as in the HCP data set, our model
lets us remove the subject variability of each seed's tractogram by calculating
the expected value across subjects:
\begin{equation}
    \label{eq:expected_subject}
    E_s(g(\rv T_{sp})) = E_s(\beta_{c} + \rv \epsilon_{c} + \rv \epsilon_{s}),
    = \beta_{c} + \rv \epsilon_{c} + E_s(\rv \epsilon_{s})
    = \beta_{c} + \rv \epsilon_{c}.
\end{equation}
where the last equality is due to $E_s(\rv \epsilon_s)=0$
(Eq. \ref{eq:ran_eff_model}). Since in our model the variabilities are normally
distributed (Eq. \ref{eq:ran_eff_model}), we can estimate the expected value across
subjects by averaging a seed's tractograms across subjects. This allows us to create
population-representative tractograms for each seed free of across-subject 
variability, which then can be clustered to create a groupwise parcellation.
\section{Experiments and Results}
In the previous section we presented a model for the cortical extrinsic 
connectivity and a clustering technique to parcellate the whole brain. Our technique
allows us to create single subject and groupwise parcellations, encoded with
different levels of granularity in a dendrogram. Now, we show the results of
applying our technique over the HCP dataset. First, we explain how the 
preprocessing step of tractography was made. Then, we elaborate 
in detail how we applied our technique. Later, we show that our groupwise 
technique creates results consistent when parceling different groups. Also,
we show that our techniques creates parcels in accordance with those by 
\citet{ThiebautdeSchotten2016} when parceling only the frontal lobe. Then,
we present a proof-of-principle that our parcels are related to brain anatomy
and functional specialization. Most of the results in this section are focused
in the groupwise case, for further information on the single-subject technique
please refer to \citet{Gallardo2016}. Finally, we study the (dis)similarity
between our groupwise parcellation and that of \citet{Glasser2016}.
\subsection{Data and Preprocessing}
\subsubsection{Human Connectome Project Dataset}
A total of 138 subjects (65 males and 73 females, ages 31-35) were randomly 
selected from the group S500 of the Human Connectome Project (HCP). For
information on the acquisition protocols please refer to \citet{VanEssen2012}.
Every subject has been already preprocessed with the HCP minimum 
pipeline~\citep{Glasser2013}. Also, each subject's cortical surface is
coregistered and represented as a triangular mesh of approximately 32000
vertices per hemisphere~\citep{Glasser2013}. For each
vertex, the corresponding label from Desikan's Atlas is known~\citep{Desikan2006}.
Finally, the group S500 contains tfMRI information representing the average
response to functional stimuli in 100 unrelated subjects (U100)\citep{Barch2013}.

\subsubsection{Probabilistic Tractography}
To create the tractograms of each subject, we performed Constrained Spherical 
Deconvolution (CSD) based tractography~\citep{Tournier2004} from a dense set of
points in the cortex. Specifically, since each subject has a mesh representing
their gray-matter/white-matter interface~\citep{Glasser2013}, we used their
vertices as seeds to create tractograms. Vertices corresponding to the medial
wall were excluded. To avoid superficial cortico-cortical fibers~\citep{Reveley2015},
we shrank each of the 138 surfaces $2mm$ into the white matter. For each subject,
we fitted a CSD model~\citep{Tournier2004} to their diffusion data using Dipy
(version 0.11)~\citep{Garyfallidis2014} and created 5000 streamlines per seed-voxel
using the implementation of probabilistic tractography in Dipy. Later, we
created a tractogram as in (Eq. \ref{eq:tractogram}) by calculating for each
seed the fraction of they particles that visited other seed-voxel.
\subsection{Parceling Subjects From the Human Connectome Project}
\begin{figure}
    \includegraphics[width=\textwidth]{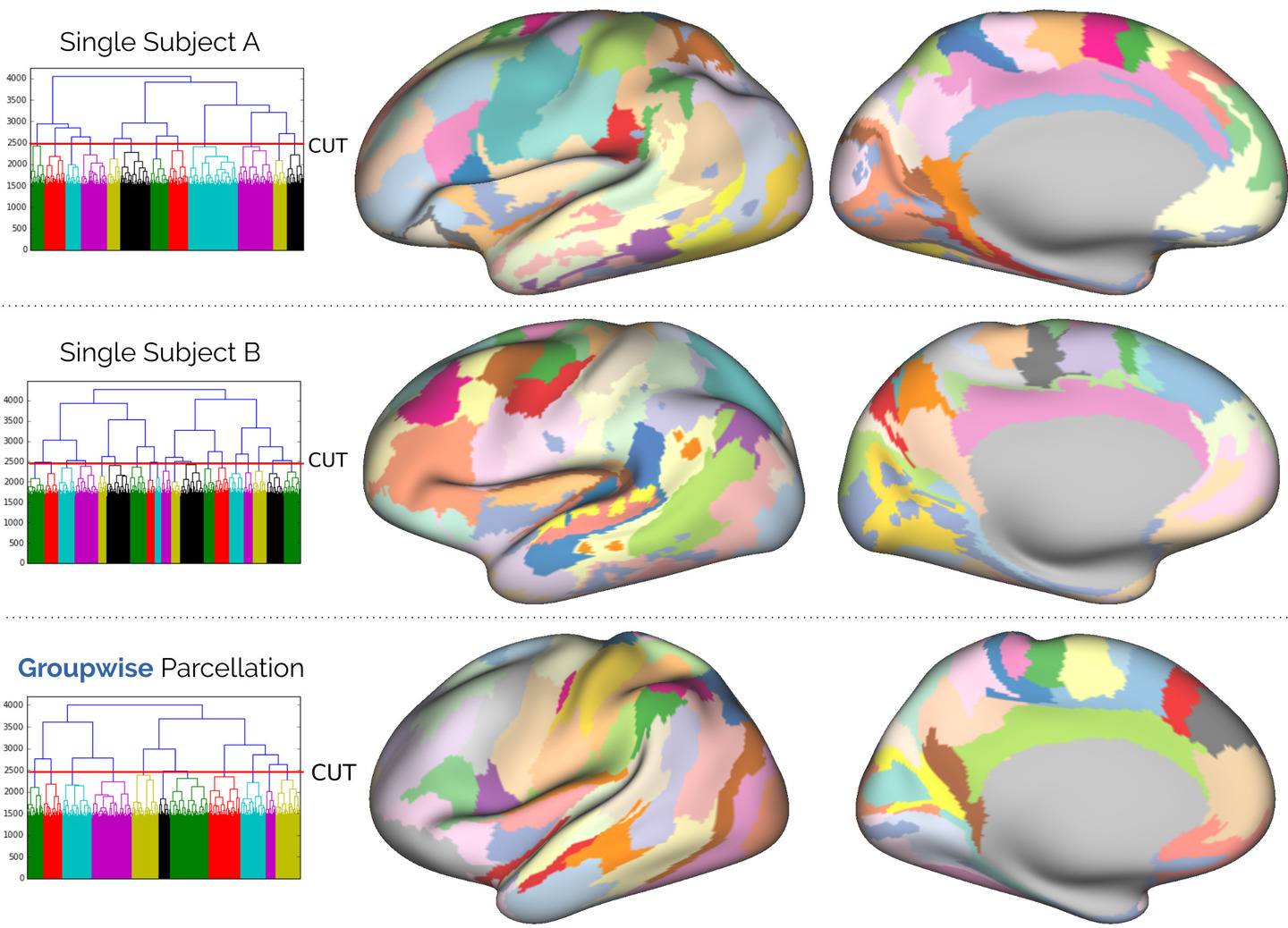}
    \caption{Examples of two single-subject parcellations and the groupwise parcellations
    		 computed with our technique. All the parcellations shown have 55 parcels.
             The corresponding dendrogram for each case, along with the chosen cut height
             (red line) are shown. The groupwise parcellation
             is based on 138 subjects from the Human Connectome Project.}
    \label{fig:single_and_group}
\end{figure}

After performing tractography, we applied our parceling technique over each
subject in our HCP sample. Specifically, we first transformed each tractogram 
with the logit function as in eq. \ref{eq:not_abuse}. Then,
we clustered the tractograms of each subject using the modified AHC algorithm
while imposing a minimum cluster size of $3mm^2$ in the finest granularity.
Two examples of obtained single-subject parcellations at a granularity of 55 
parcels are shown in fig. \ref{fig:single_and_group}.
To create the groupwise parcellation, we took advantage of the vertex
correspondence across subjects in the HCP data set~\citep{Glasser2013}. After
transforming the tractograms with the logit transform, we computed the average
connectivity of each seed by averaging its tractograms across-subject. Then, we
computed the groupwise parcellation by clustering the averaged tractograms with our
proposed technique (sec. \ref{sec:parceling_methodologies}). The obtained groupwise 
parcellation at a granularity of 55 parcels is shown in fig.
\ref{fig:single_and_group}.
\subsection{Groupwise Parcellation Technique Consistency}
\begin{figure}
    \includegraphics[width=\textwidth]{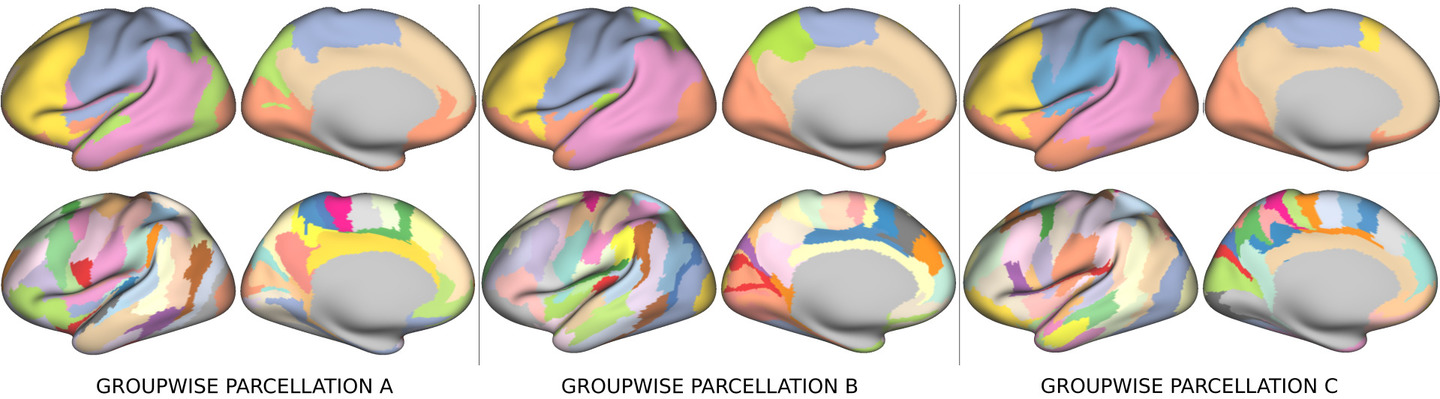}
    \caption{Groupwise parcellations of 3 disjoint groups of 46 people each.
             We show results from the same dendrogram cut to get 6 parcels (upper)
             and 55 parcels (lower). Labels with best overlap in upper figures share
             the same color. Notice that there are two different shades of blue for 
             the group C.}
    \label{fig:groups}
\end{figure}

\begin{figure}
    \includegraphics[width=\textwidth]{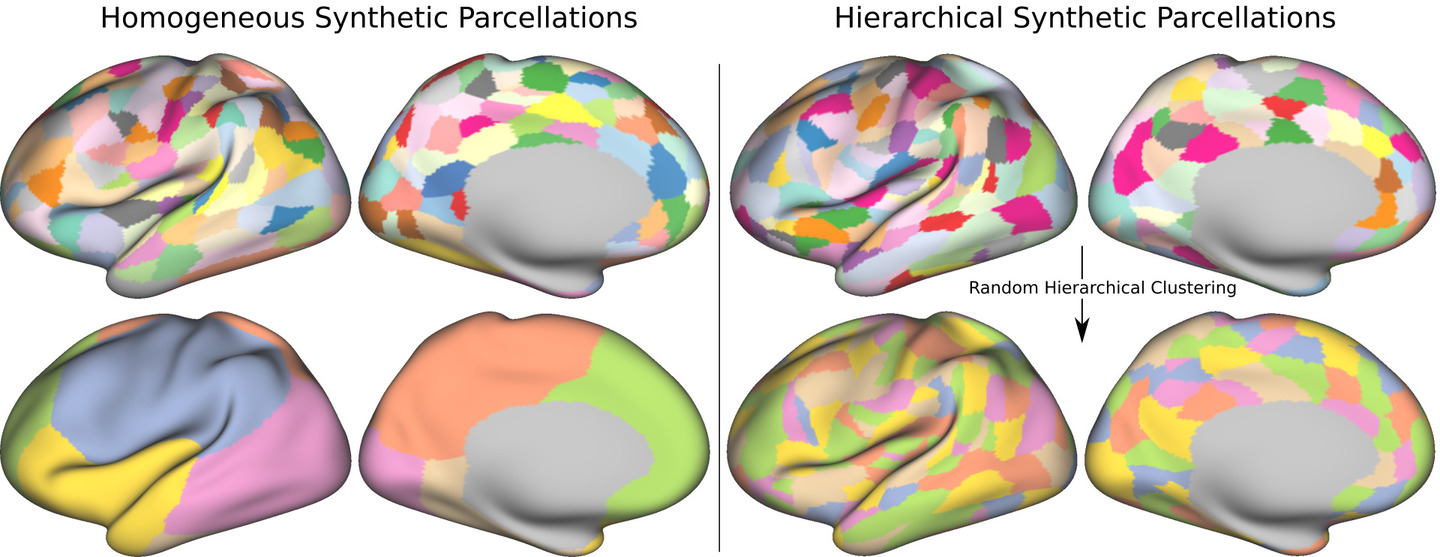}
    \caption{Examples of synthetic parcellations created to compute a baseline
             adjusted rand index. Parcellations on the left were created by 
             dividing the brain in a homogeneous way, inspired by the random
             parcellation presented in \citet{Paristot2015}. Parcellations on
             the right were created by randomly merging parcels of a coarse
             parcellation.}
    \label{fig:fake_parcels}
\end{figure}

\begin{figure}
    \centering
    \includegraphics[width=\textwidth]{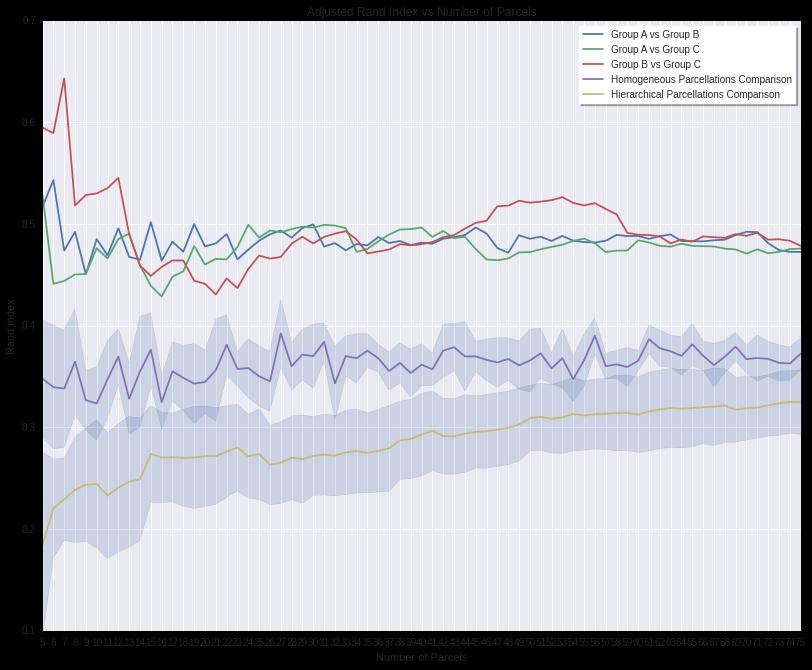}
    \caption{Adjusted Rand Index obtained when comparing: (red) Group A vs Group B;
             (blue) Group A vs Group C; (green) Group B vs Group C; (purple)
             Synthetic Homogeneous Parcels and (yellow) Synthetic hierarchical
             Parcels.}
    \label{fig:real_vs_fake}
\end{figure}

To study the consistency of our technique, we randomly divided our HCP subject
sample 
in 3 disjoint groups, trying to maintain the same proportion of males and females
on each. The resulting groups had: 24 females, 22 males (group A); 23 females, 
23 males (group B) and 28 females, 18 males (group C). For each group we computed
their groupwise parcellation. The resulting parcellations at two different levels 
of granularity are shown in fig. \ref{fig:groups}.
To study the similarity between the obtained groupwise parcellations, we compared
them at different levels of granularity using the adjusted Rand
index~\citep{Hubert1985}. To have a baseline for the comparisons, we generated
random parcellations of the cortex and computed the similarity between them. 
We computed two types of random parcellations: The first one is an homogeneous
random parcellation with $n$ parcels, inspired in a method used by \citet{Paristot2015}. 
To compute it, we start by choosing $n$ starting points in the cortex, then, we
randomly expand each parcel
on the cortex. By comparing these random parcellations between them we compute 
the minimum obtainable Rand index by mere chance at each level of granularity. 
In the second type of random parcellation, 
we simulate the behavior of our technique. For this, we create a parcellation 
with $300$ parcels and then, we iteratively merge two parcels chosen at random until
all the parcels are merged in one. By comparing these random parcellations between
them we
obtain the minimum obtainable Rand index by a random Hierarchical Clustering
Algorithm. 
Examples of these random parcellations can be seen in Fig \ref{fig:fake_parcels}. 
The baselines presented in fig. \ref{fig:real_vs_fake} (yellow and violet lines)
were computed by comparing $1000$ of these random parcels at different levels
of granularity.
The result of comparing the groupwise parcellations of each group appear in
fig. \ref{fig:real_vs_fake}. The figure shows that the similarity between
our groupwise parcellations (lines red, green and blue) are significantly
higher than the baselines (violet and yellow). That is, the similarity
between our parcellations differs (for most cases) more than 3 standard 
deviations from the baselines' mean. Moreover, the similarity between our
results differs more
than 4 standard deviations from the comparison between synthetic hierarchical
parcels. This results show that our groupwise parceling technique creates
consistent parcellations.
\subsection{Relationship with a frontal lobe parcellation}
Here we assess the agreement of our technique with an state-of-the-art extrinsic 
connectivity parceling technique. We do so by using our technique to parcellate 
the frontal lobe and compare our result against that of \citet{ThiebautdeSchotten2016}. 
In their work, \citet{ThiebautdeSchotten2016} use a principal component analysis (PCA)
statistical framework to parcellate the frontal lobe. They obtain a parcellation
with 12 parcels. Then, they show that each one of these parcels possess a functional
specialization by using the Decode tool\footnote{http://www.neurosynth.org/decode/}
from Neurosynth \citep{Yarkoni2011}. 
Thiebaut's parcellation is currently available in Neurovault \citep{Gorgolewski2016}
as an annotated volume\footnote{http://neurovault.org/collections/1597/}, registered
on the Colin27 template \citep{Holmes1996}. We downloaded this
parcellation and projected its parcels into a dense mesh representing the
cortex of the Colin27 template. The dense mesh had the same amount of vertices
as our chosen HCP subjects, and such vertices were coregistered with the
HCP subjects' cortical surfaces ones.

\begin{table}[t]
\centering
\scalebox{0.9}{
  \begin{tabular}{@{}cccc@{}}
      \multicolumn{4}{c}{\textbf{Table 1. Correlation value reported (Neurosynth)}}  \\ \midrule
\textbf{Parcel} & \textbf{Term} & \textbf{$r$ (Thiebaut et al.)} & \textbf{$r$ (Ours)}\\
\textbf{1} & foot  & 0.267 & \textbf{0.319} \\
\textbf{2} & motor & 0.129 & \textbf{0.208} \\
\textbf{3} & eye field & 0.081& 0.048\\
\textbf{4} & speech production &0.077&\textbf{0.138}\\
\textbf{5} & pre sma &0.245&0.234\\
\textbf{6} & phonological &0.206&0.019\\
\textbf{7} & - &-&-\\
\textbf{8} & executive control & 0.049 & 0.042\\
\textbf{9} & - &-&-\\
\textbf{10}& semantic &0.178&\textbf{0.226}\\
\textbf{11}& social &0.137&0.110\\
\textbf{12}& semantic &0.139&0.086\\      \bottomrule
\end{tabular}}
\vspace{0.3cm}
\caption*{Table 1. Spatial correlation value reported by Neurosynth for specific
                   terms in each parcel of \citet{ThiebautdeSchotten2016} and
                   for our parcels. Enumeration comes from figure~\ref{fig:frontal}.} 
\end{table}

\begin{figure}
    \includegraphics[width=\textwidth]{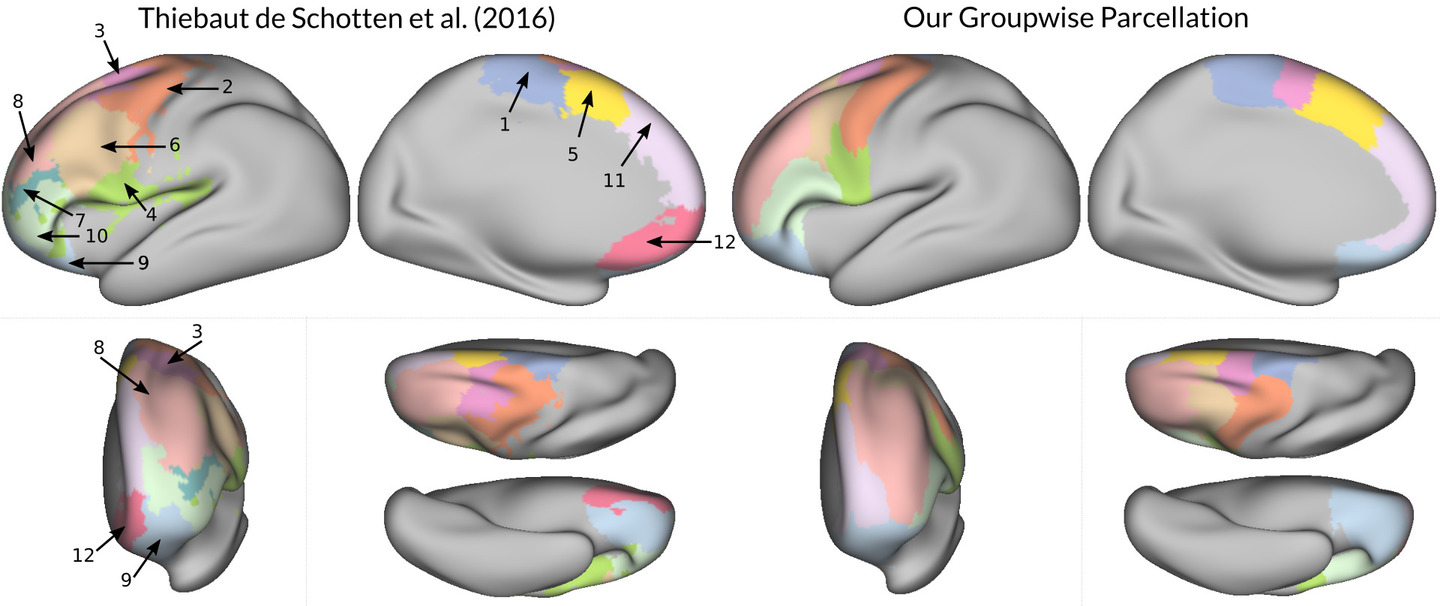}
    \caption{\citet{ThiebautdeSchotten2016} parcellation (left) and our groupwise
    	     parcellation using only tractograms from the frontal lobe (right).
             Our parcels are colored after the parcel from \citet{ThiebautdeSchotten2016}
             with which they best overlap.}
    \label{fig:frontal}
\end{figure}

From the Desikan Atlas \citep{Desikan2006} of each of our HCP subjects, we
derived a groupwise mask for the frontal lobe. Then, we computed a groupwise 
parcellation with our technique, using only the tractograms in the mask.
Figure \ref{fig:frontal} shows both the parcellation downloaded from Neurovault
and our groupwise parcellation projected in the Colin template cortical surface.
The figure shows our parcellation with 10 parcels since this level of
granularity showed the best Rand index against the Thiebaut's parcellation.
The colors of each parcel in our groupwise parcellation were picked in base to
the position and amount of overlapping with the Thiebaut's parcels on the
surface. While the similarity according to the Rand index is not significantly
high ($0.4$), some visual similarity can be observed on the obtained parcellation,
particularly in the blue, yellow, orange and green parcels. Moreover, as shown in
table 1, our parcels show the same or even a higher level of functional
specialization when processed with Neurosynth.

\begin{figure}
    \centering
    \includegraphics[width=\textwidth]{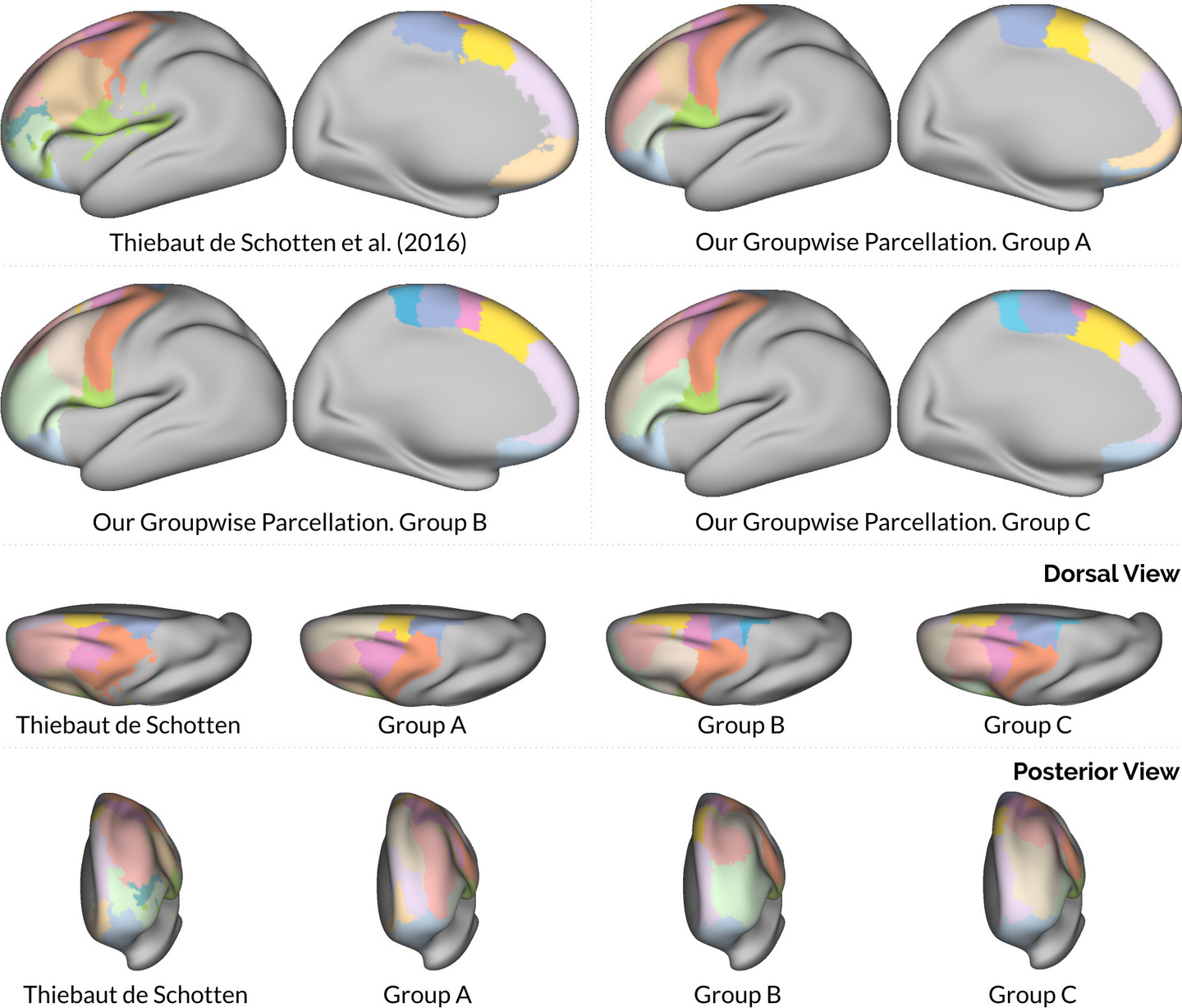}
    \caption{\citet{ThiebautdeSchotten2016} parcellation (top-left) and our
    		 frontal lobe groupwise parcellations computed over 3 disjoint 
             groups of subjects. Our parcels are colored after the parcel from
             \citet{ThiebautdeSchotten2016} with which they best overlap.}
    \label{fig:indices_by_lobe}
\end{figure}

To study the consistency of our result we computed the frontal lobe groupwise
parcellation in each of the 3 disjoint groups from the previous experiment. Figure~
\ref{fig:indices_by_lobe} shows the three obtained parcellation alongside
the Thiebaut's one. The obtained parcels show consistency, obtaining
an adjusted Rand index score of $0.61 \rpm 0.05$ between them.

\subsection{Anatomical Relationship and Functional Specialization of Our Parcels}
Here we present a proof of concept that our technique creates parcels within
anatomical boundaries and with functional meaning. To do so, first, we extracted
a parcellation with 55 parcels from the groupwise parcellation
computed from the 138 subjects. This was made to get a  parcellation with coarse
granularity while having at least the amount of 
parcels in the anatomical atlas of Desikan~\citep{Desikan2006} (36 parcels). We
compare this extracted parcellation against the Desikan Atlas and a functional
study made to every subject in the HCP~\citep{Glasser2013}.
\subsubsection{Relationship with Anatomical Boundaries}
\begin{figure}
    \includegraphics[width=\textwidth]{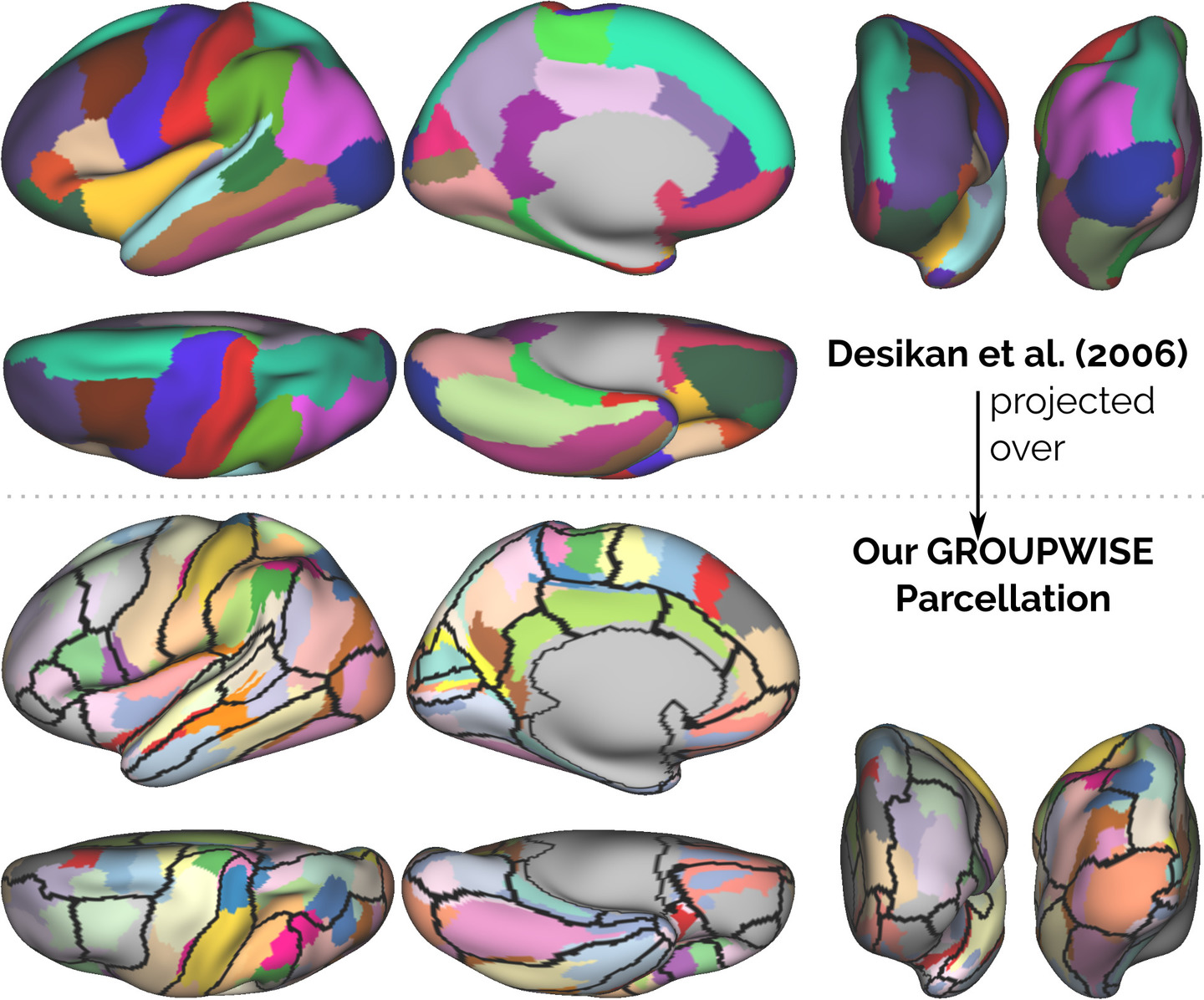}
    \caption{Relation between our pure extrinsic parcellation and the anatomical
             atlas of Desikan~\citep{Desikan2006}. Desikan atlas
             projected over the groupwise parcellation with 55 parcels.
             Insula; Cingulate; Lateral-Occipital; Fusiform; Superior Frontal;
             Lingual; Sensory and  Motor Cortex appear to be found.}
    \label{fig:anatomical_zoom}
\end{figure}
To assess if some anatomical structures were present in the dendrogram and 
if our resulting parcels were subdividing them, we compared our extracted
parcellation with the Desikan atlas~\citep{Desikan2006}. To do so, we projected 
the Desikan regions over our parcels and then calculated: how many of our parcels
were contained by a anatomical region in more than a $90\%$, and which anatomical
regions were contained inside of one of our parcels. Using this criterion, the Insula; 
Cingulate; Lateral-Occipital; Fusiform; Superior Frontal; Lingual; Sensory and
Motor Cortex appear to be found as shown in Fig. \ref{fig:anatomical_zoom}.
\subsubsection{Functional Specialization.}
\begin{figure}
    \includegraphics[width=\textwidth]{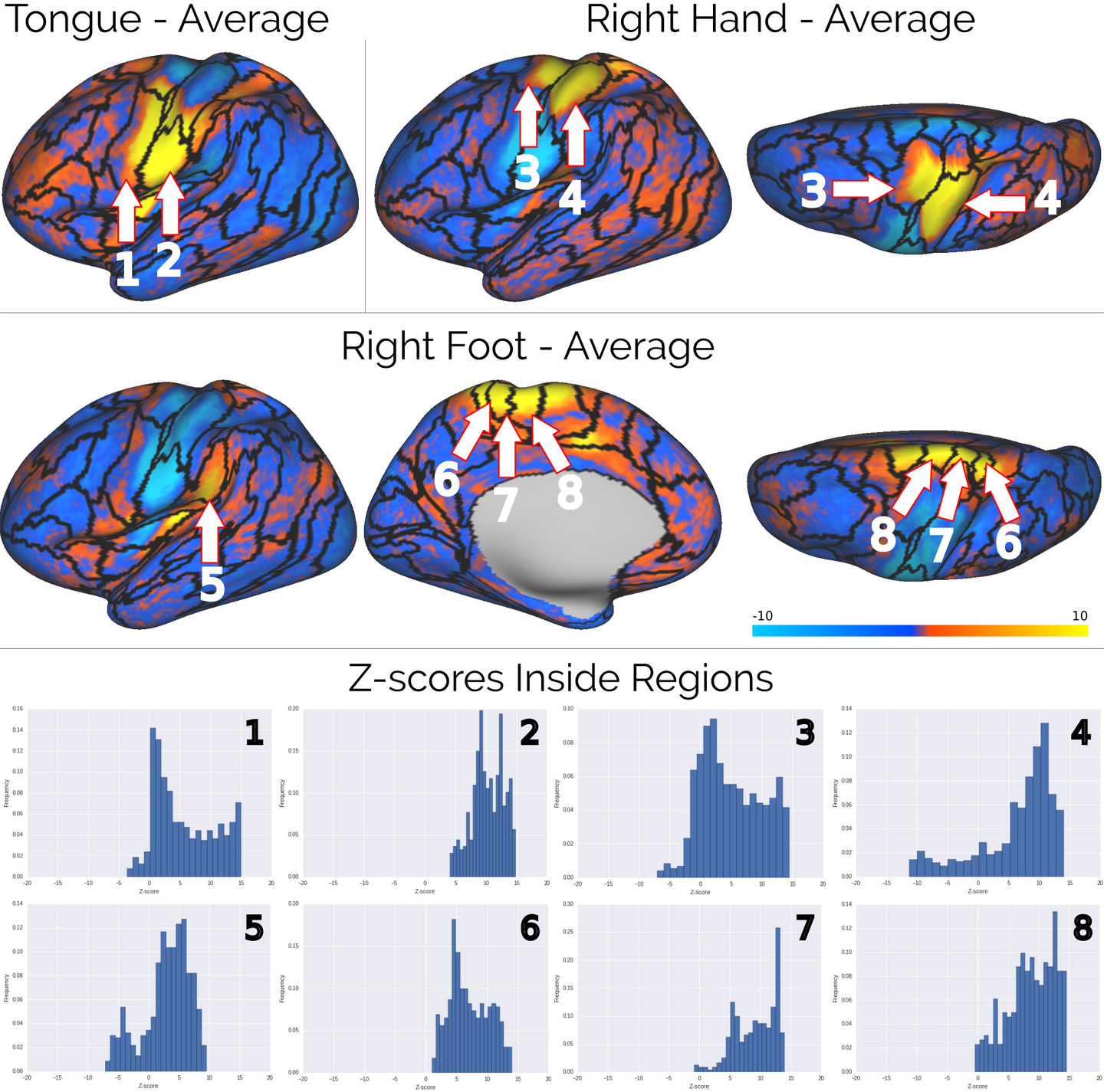}
    \caption{Our groupwise parcellation with 55 parcels projected over
             z-scores representing responses to motor tasks. Each histogram
             shows the distribution of z-score inside our parcels. The null or
             small fraction of negative values shows the functional specialization
             of our parcels}
    \label{fig:function_motor}
\end{figure}

\begin{figure}
    \includegraphics[width=\textwidth]{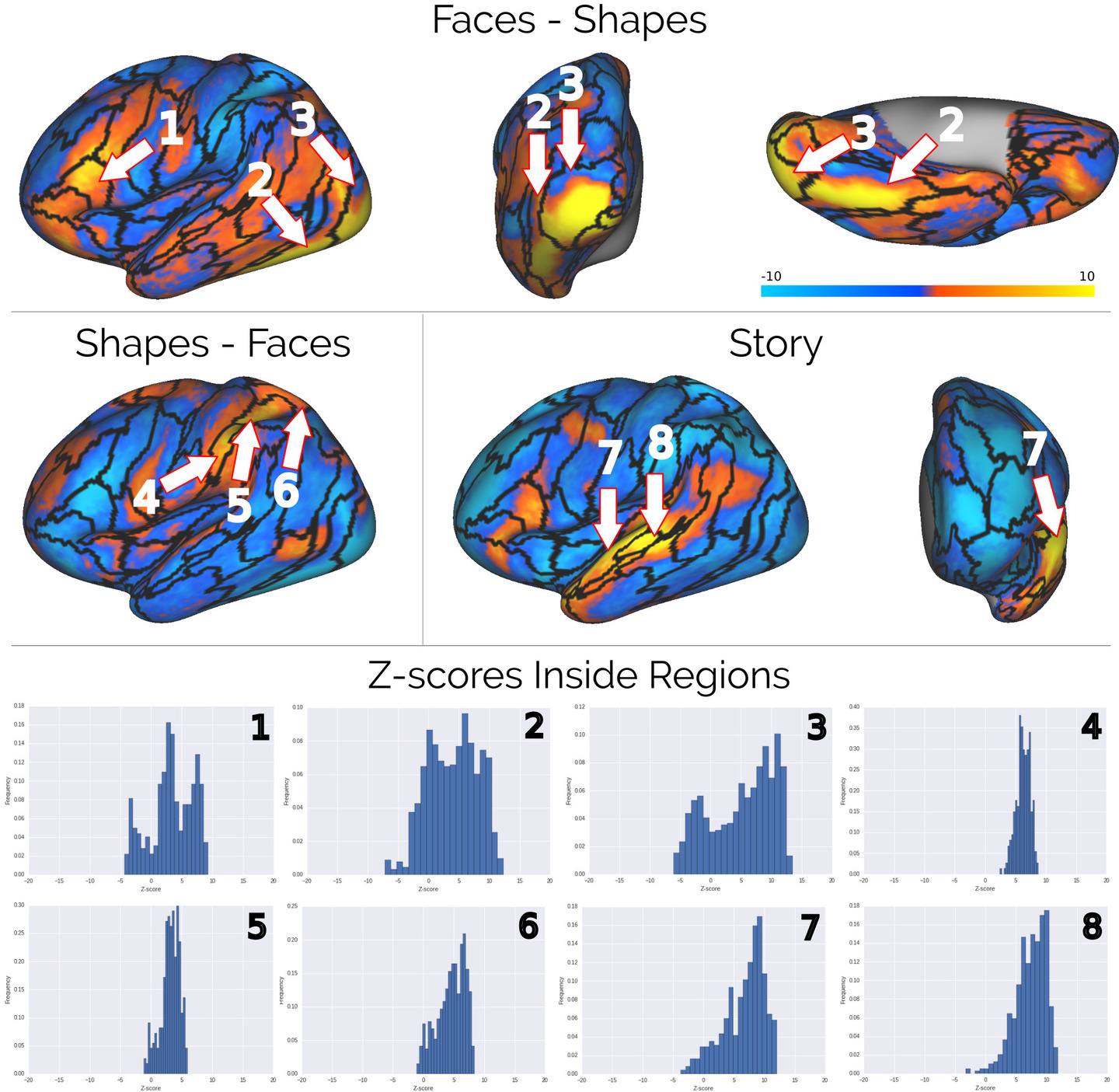}
    \caption{{Our groupwise parcellation with 55 parcels projected over
             z-scores representing responses to cognitive tasks. Each histogram
             shows the distribution of z-score inside our parcels. The null or
             small fraction of negative values shows the functional specialization
             of our parcels}
}
    \label{fig:function_cognition}
\end{figure}

To study the relationship between our parcels and brain function, we projected our 
parcels over z-score maps representing responses to functional 
stimuli~\citep{Barch2013}. These maps are available as part of the HCP data, 
and represent the average activation of 100 subjects. In particular, we used
the maps related to 
the following tasks: right hand, foot and tongue movement; face, shape
recognition  and story categorization. For information on the functional tasks,
acquisition and processing of this data please refer to \citet{Barch2013}. 
Figure \ref{fig:function_motor} shows our parcels projected over contrasts
in motor tasks. In particular, our parcels are projected over the following
contrasts: tongue-average; hand movement-average and foot movement-average. 
Figure \ref{fig:function_cognition} shows our parcels projected over contrasts
in cognitive tasks: face-shape recognition; shape-face recognition and short-story
categorization. The figures show a good overlap between our parcels and the
regions with maximum activation of each task. In both figures the distribution
of z-scores inside of specific regions are shown as histograms. Further 
information about the z-score is present in tables 2 and 3. These tables show
that our parcels contain zero or few negatives values; that the mean of their
contained z-score is always positive and also, that many of those parcels 
enclose the maximum achievable z-score.

\begin{table}[t]
\centering
\scalebox{0.9}{
  \begin{tabular}{@{}cccccc@{}}
      \multicolumn{6}{c}{\textbf{Table 2. Statistics on z-score distribution
                                 in parcels from figure
                                 \ref{fig:function_motor}}}  \\ \midrule
\textbf{Contrast} & \textbf{Parcel} & \textbf{Min.} & \textbf{Max.} & \textbf{Mean $\rpm$ Std. Dev.} & \textbf{Max. Score in Map}\\
T-Avg & \textbf{1} & -3.62  & 15.03 & $5.67 \rpm 4.91$ & 15.03 \\
T-Avg & \textbf{2} & 4.11 & 14.88 & 10.30 $\rpm$ 2.56 & 15.03 \\
RH-Avg & \textbf{3} & -7.02  & 14.50 & 5.05  $\rpm$ 4.95 & 14.50 \\
RH-Avg & \textbf{4} & -11.25  & 14.07  & 6.35 $\rpm$ 6.25 & 14.50\\
RF-Avg & \textbf{5} & -7.10 & 9.57& 2.99 $\rpm$ 3.84 & 14.56 \\
RF-Avg & \textbf{6} & 1.04& 14.01 & 7.13 $\rpm$ 3.20 & 14.56 \\
RF-Avg & \textbf{7} &-0.83 & 13.98& 9.23 $\rpm$ 3.32 & 14.56\\
RF-Avg & \textbf{8} &-0.46 & 14.56& 8.73 $\rpm$ 3.81 & 14.56 \\ \bottomrule
    \end{tabular}}
\vspace{0.3cm}
\caption*{Table 2. Minimum; maximum and mean z-score contained by each of the
          parcels enumerated in figure \ref{fig:function_motor}. The highest 
          z-score of each map is reported to facilitate comparison. T-Avg: Tongue
          movement versus average; RH-Avg: Right Hand Movement versus average; RF-Avg: 
          Right Foot Movement versus average.}
\end{table}

\begin{table}[t]
\centering
\scalebox{0.9}{
  \begin{tabular}{@{}cccccc@{}}
      \multicolumn{6}{c}{\textbf{Table 3. Statistics on z-score distribution
      in parcels from figure \ref{fig:function_cognition}}} \\ \midrule
\textbf{Contrast} & \textbf{Parcel} & \textbf{Min.} & \textbf{Max.} & \textbf{Mean $\rpm$ Std. Dev.} & \textbf{Max. Score in map}\\
Faces-Shapes & \textbf{1} & -4.33 &  9.28 & 3.35 $\rpm$ 3.51 & 13.45 \\
Faces-Shapes & \textbf{2} & -7.16 & 12.36 & 4.01 $\rpm$ 4.09 & 13.45 \\
Faces-Shapes & \textbf{3} & -6.07 & 13.45 & 5.16 $\rpm$ 5.25 & 13.45 \\
Shapes-Faces & \textbf{4} & -5.73 &  5.37 & 0.93 $\rpm$ 1.78 &  8.79 \\
Shapes-Faces & \textbf{5} & -4.11 &  7.67 & 1.11 $\rpm$ 2.11 &  8.79\\
Shapes-Faces & \textbf{6} & -1.13 &  5.94 & 3.17 $\rpm$ 1.49 &  8.79 \\
Story & \textbf{7} & -3.72 & 12.02 & 6.72 $\rpm$ 3.35 & 12.02\\
Story & \textbf{8} & -3.24 & 11.92 & 7.41 $\rpm$ 2.50 & 12.02\\ \bottomrule
\end{tabular}}
\vspace{0.3cm}
\caption*{Table 3. Minimum; maximum and mean z-score contained by each of the 
          parcels enumerated in figure \ref{fig:function_cognition}. The
          highest z-score of each map is reported to facilitate comparison. Faces-Shapes: 
          Face recognition versus shape recognition contrast; Shapes-Faces: Shape recognition
          versus face recognition; Story: Short story categorization.}
\end{table}
\subsection{Relationship With a Multi-Modal Parcellation of the Cortex}
Finally, we study the (dis)similarities between our groupwise parcellation and that
of \citet{Glasser2016}. In their work, \citet{Glasser2016} compute a parcellation
of the whole cortex using information from different MRI modalities. In particular,
they use information from task functional MRI; resting state functional MRI;
myelin maps computed from T1 and T2 images and cortical thickness. It is important
to remark that dMRI data, in which our work is solely based, was not used to construct 
their parcellation.

\begin{figure}
    \centering
    \includegraphics[width=\textwidth]{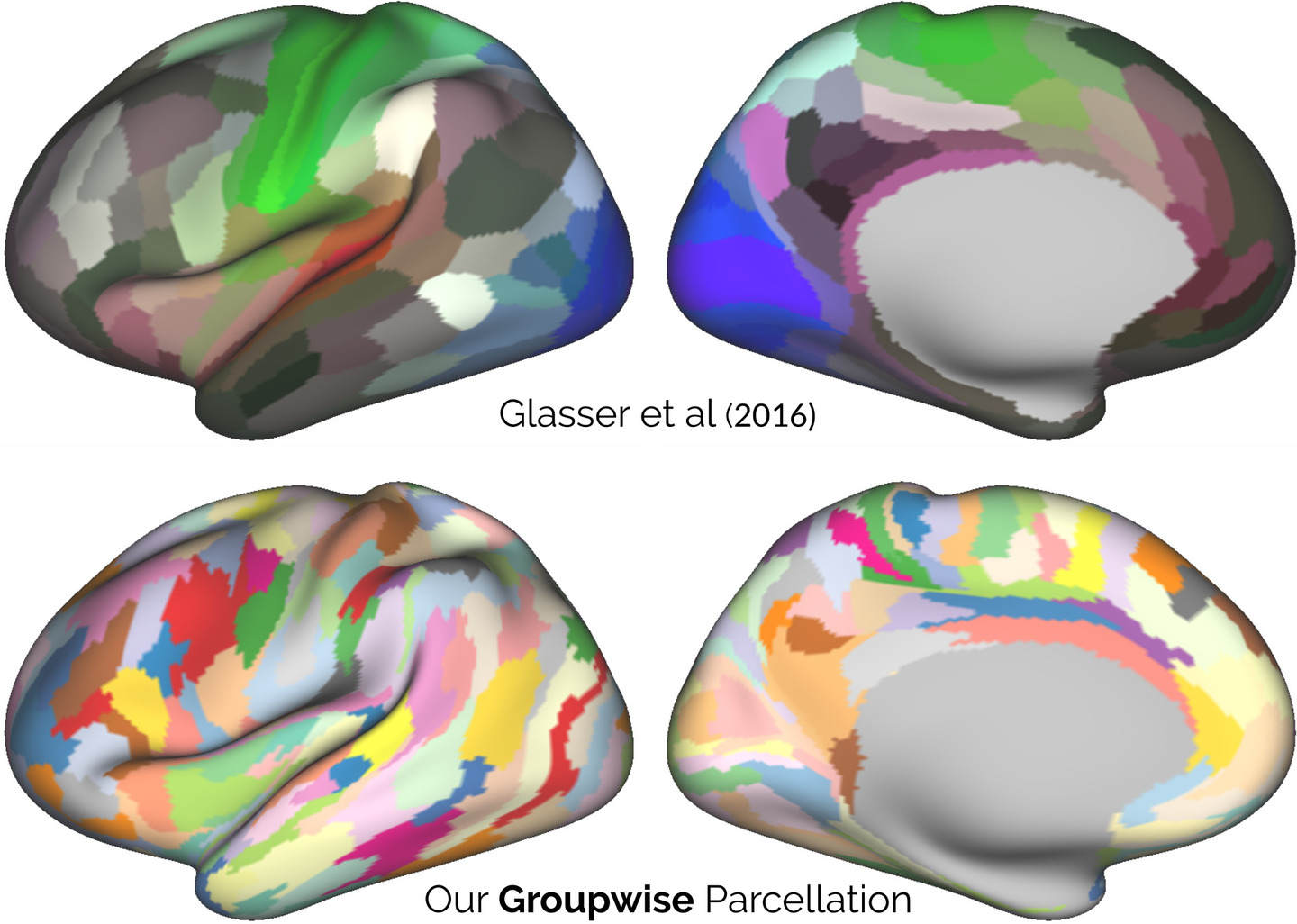}
    \caption{\citet{Glasser2016} parcellation (upper) and our
    		 groupwise parcellations computed from 138 HCP subjects. Both 
             parcellations contain 180 parcels. There's almost no overlap according
             to the adjusted Rand index between them (0.28).}
    \label{fig:glasser_and_me}
\end{figure}

\begin{figure}
    \centering
    \includegraphics[width=\textwidth]{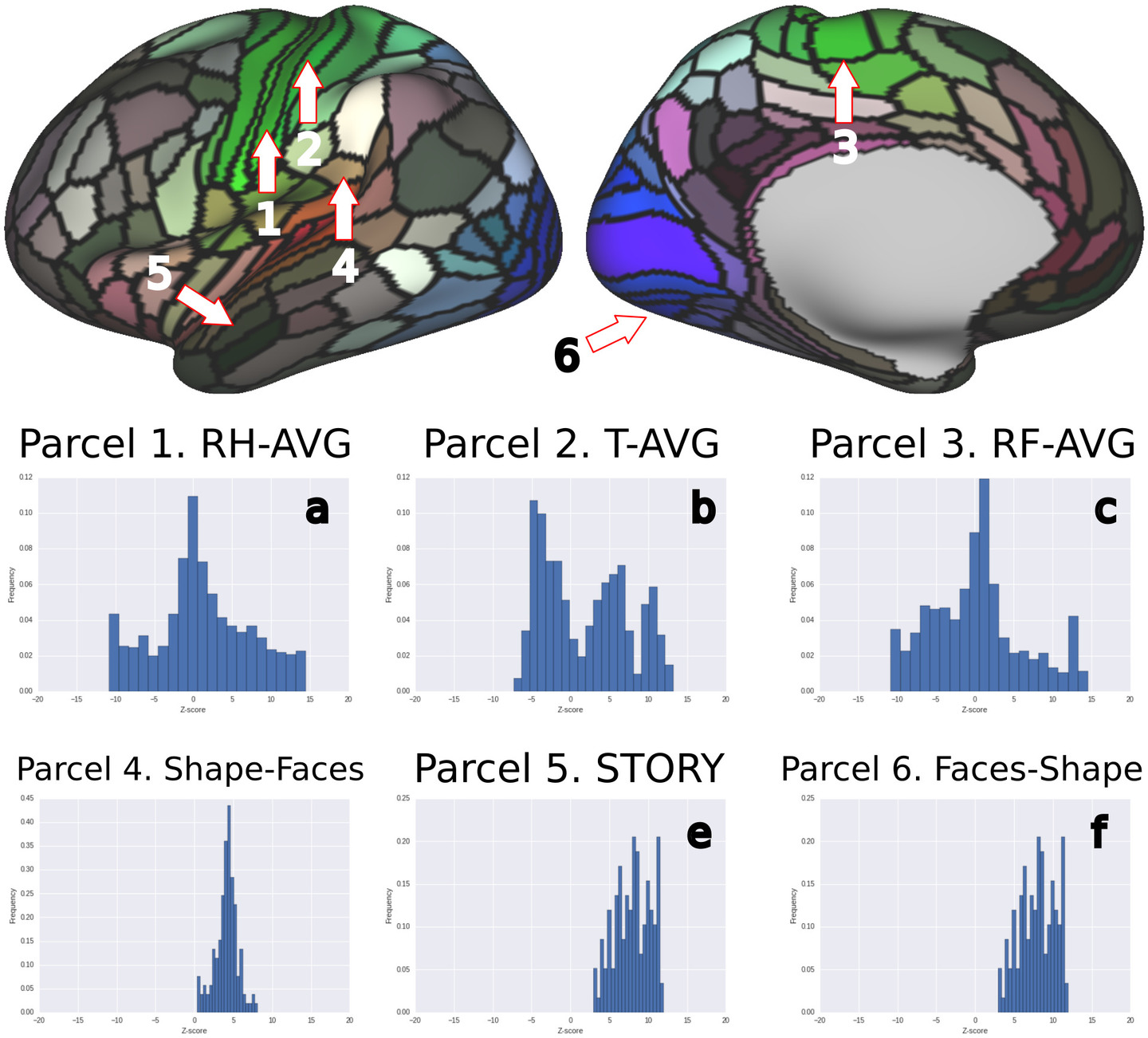}
    \caption{\citet{Glasser2016} parcellation (upper) and histograms of 
             z-score contained in different parcels for different functional
             task. (a) Histogram for parcel 1 for the contrast related to
             Tongue movement. (b) Histogram for parcel 2 for the contrast
             related to Tongue movement. (c) Histogram for parcel 3 for the
             contrast related to Right Foot movement. (d) Histogram for parcel
             4 for the contrast Shape recognition vs Face recognition. (e)
             Histogram for parcel 5 for the contrast related to Story
             Categorization.
             (f) Histogram for parcel 5 for the contrast Face recognition vs Shape
             recognition. The histograms (d); (e) and (f) correspond to the parcels
             with the greatest mean z-score of their respective tasks.}
    \label{fig:glasser_functional}
\end{figure}

To compare our results against Glasser's atlas, we first extracted a parcellation 
of 180 parcels from the groupwise dendrogram of our 138 HCP subjects. That is, we 
extracted a parcellation with the same number of parcels as Glasser's one. Figure
\ref{fig:glasser_and_me} show both parcellations side by side. 
We compared both parcellations using the adjusted Rand Index, obtaining a score
of 0.28. Such low score indicates that there's almost no similarity between our
result and that of \citet{Glasser2016}. Also, there's no relationship with
our groupwise parcellation with 55 parcels used in the previous section since
Glasser's parcels (finest) do not subdivide ours (coarsest).
Since Glasser's parcellation comes from functional information in the HCP, we
studied the functional specialization of its parcels in the same manner as 
previous section. Figure \ref{fig:glasser_functional} shows the histogram of
z-score contained for some parcels when using the same maps as in section
Functional Activations. It's important to remark that the z-score maps used
come from responses to functional stimuli of HCP subjects \citep{Glasser2013}.
In particular, histograms a; b and c in fig. \ref{fig:glasser_functional} show
that their subdivisions of the sensori-motor cortex contain a wide range of 
z-scores, centered in zero.

\section{Discussion}
In this work we presented a parsimonious statistical model for long-ranged
axonal connectivity. Our model (section \ref{sec:cortical_model}), assumes that
the cortex is divided in patches of homogeneous extrinsic connectivity, as histological
results showed in the macaque brain \citep{Schmahmann2006}. By borrowing ideas
from statistical clustered data models~\citep{Pendergast1996}, our model accounts
for the variability in the axonal connections of a patch's neurons and for 
variability in patch boundaries across subjects.

Taking advantage of our proposed model, in Section \ref{sec:parceling_methodologies}
we presented an efficient
technique to parcellate the cortex based on its extrinsic connectivity. Our
technique uses only dMRI information, without the need of relying on
initial parcellations~\citep{Clarkson2010}. Also, our technique allows 
parcellation of the whole cortex, overcoming the problem of working with only part
of it~\citep{Lefranc2016, Roca2009, ThiebautdeSchotten2014, ThiebautdeSchotten2016}.
Additionally, our technique allows creation of both single subject and
groupwise parcellations independently, avoiding the need to impose
constraints between them~\citep{Clarkson2010, Roca2010, Paristot2015}.

Inspired by \citet{Moreno-Dominguez2014}, our technique uses Hierarchical
Clustering to comprise multiple granularities of the same parcellation in a
dendrogram. The fundamental difference is how we compare and merge tractograms 
during the clustering process. \citet{Moreno-Dominguez2014} use Centroid
Clustering~\citep{Murtagh1985} with the cosine distance. This can lead to an
erroneous parcellation since the centroid criterion doesn't minimize the cosine
distance between points. Also, this method makes dendrograms with inversions
\citep{Murtagh1985}, which are then removed heuristically. In our case,
using a Logistic Random Effect model
(eq. \ref{eq:ran_eff_model}) allowed us to transform the tractograms into a
Eucliden space (sec. \ref{sec:parceling_methodologies}).
Then, we use Ward's Hierarchical method~\citep{Murtagh2011}, which creates
clusters with minimum intra-cluster variance. We can use this algorithm since
its only hypothesis is that the features to cluster are in a Euclidean space. 
One advantage of using Ward's method is that we can use the Lance and Williams
\citep{Lance} formula to update the disimilarity between tractograms at each
iteration. The formula gives us the disimilarity between the new centroid 
created at each step and the rest of the existing tractograms in constant time.
As far
as we know there's no Lance and Williams formula when using the cosine distance
with the centroid linkage. This allows us to
lower the time complexity of our algorithm with respect to Moreno-Dominguez.
Another advantage of using Ward's clustering is that our
resulting dendrograms do not have inversions, which means that we don't need to
post-process them. As in \citet{Moreno-Dominguez2014}, we also create the
dendrogram using only one comprehensive parameter: the minimum size of each
cluster. This parameter imposes the local coherence criterion. Using Hierarchical
Clustering allows us to overcome the need of other techniques~\citep{Paristot2015}
to specify an expected number of clusters, which imposes the need to recompute
the whole pipeline each time a new parcellation is required, and, using our
Logistic Random Effects Model, allows us to use a hierarchical clustering 
algorithm which minimizes the intra-cluster variance.

\subsection{Our Groupwise Parcellations are Consistent Across Similar Groups:}
We assessed the consistency of our groupwise parcellation by quantifying the
consistency across 3 disjoint groups of 46 subjects each. The consistency is
shown by the adjusted
Rand index in Fig.~\ref{fig:real_vs_fake}, which quantifies consistency across
parcellations~\citep{Hubert1985}. As seen in Fig.~\ref{fig:real_vs_fake} whole-cortex
parcellations obtained with our method are consistent across groups, and the
Adjusted Rand Index is significantly higher, i.e.\ more than 3 standard
deviations, for all granularities when compared with the null case of
randomly-generated parcellations. 

Our whole-cortex groupwise parcellation reaches a maximum consistency score when
the cortex is divided in 6 regions, see Fig.~\ref{fig:real_vs_fake}.  As seen in
Fig. \ref{fig:groups}, these parcellations are consistent with specific
anatomo/functional networks: the frontal lobe section anterior to the prefrontal
cortex is shown in yellow; the sensorimotor area is shown in cyan, the cingulate
area is shown in beige; the fronto-occipital connection in orange, and the
temporo-parietal system in pink. 

\subsection{Our Method Creates Parcels in Agreement With a Single-Lobe Parceling
            Technique Extant in the Literature.}
We showed that our technique obtains results similar to another method extant
in the literature. We did so by parceling only the frontal and showing the visual
similarity between our resulting parcels and those obtained by
\citet{ThiebautdeSchotten2016}. Moreover, the blue, pink and green parcels
in fig. \ref{fig:frontal} share not only similar boundaries and location, but
also functional specialization (Table 1). In some cases our parcels possess even higher
spatial-correlation with functional task according to
Neurosynth's \citep{Yarkoni2011} Decode tool\footnote{http://neurosynth.org/decode/}. We assessed the 
consistency of our obtained groupwise parcellation by computing the groupwise 
frontal lobe parcellation of three disjoints groups of 46 subjects and comparing
them using the adjusted Rand index. The obtained value of $0.61$ shows that our
parcellation of the frontal lobe is consistent across groups.
\subsection{Our Method Creates Several Parcels in Agreement with Brain Anatomy.}
\begin{figure}
    \includegraphics[width=\textwidth]{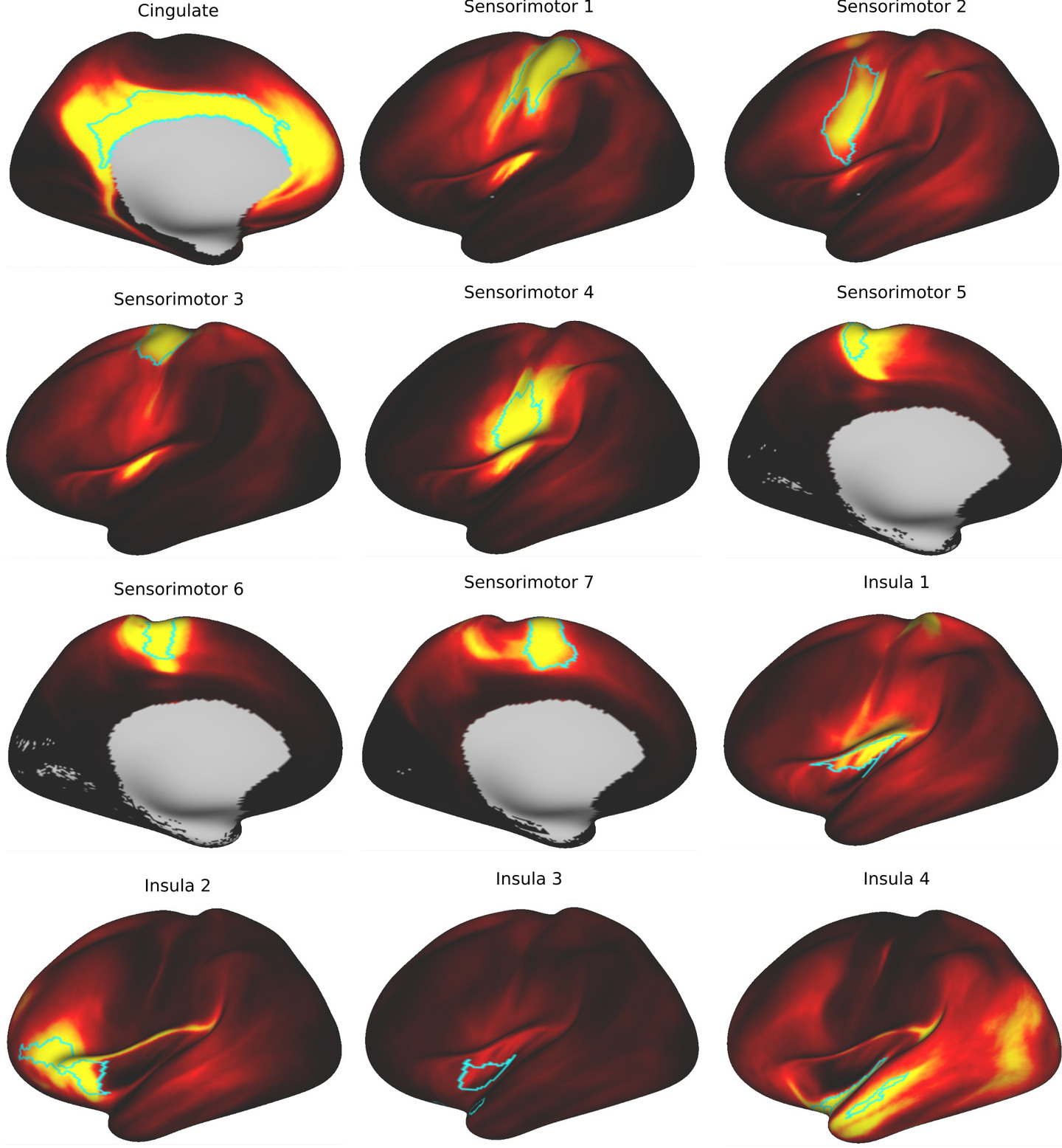}
    \caption{Connectivity fingerprint for different parcels in our groupwise
             parcellation. The names in the titles are given after the anatomical
             structure that they subdivide (or contain, as with the Fusiform). }
    \label{fig:conn_fing}
\end{figure}

We showed that many of our parcels are in agreement with brain anatomy. 
In particular, we showed that in our groupwise parcellation, with 55 parcels,
the following anatomical structures appeared to be found: Cingulate; Insula;
Lateral-Occipital; Fusiform; Superior Frontal; Lingual; Motor and Sensory cortex.
Here we discuss why some of these parcels were found and how are their 
conectivity fingerprints.
In the case of the Cingulate, its fingerprint, shown in fig. \ref{fig:conn_fing}, is
strongly related with the Cingulate Fascicle (CF) pathway. This
is consistent with the 
fact that the seeds located in the Cingulate will end up into the CF after being 
pushed in the white-matter. In the case of the Insula, each subdivision
showed a specific pattern of connectivity as shown in fig. \ref{fig:conn_fing}. 
These parcels show a gradient of connections from the occipital lobe to the frontal
lobe consistent with that of \citet{Ghaziri2015}. In the
Lateral-Occipital region, we see a specific pattern of local connectivity which
cannot be attributed to gyral bias since the Lateral-Occipital covers many sulci
and gyrus. In the case of the fusiform, it is almost completely contained in one
of our parcellations, which goes from the Fusiform up to the Lateral-Occipital
(fig. \ref{fig:anatomical_zoom}). 
This could add evidence to the hypothesis that the Fusiform plays a role in visual
tasks \citep{Kanwisher2006, Yeatman2014}. Finally, the Motor and Sensory cortex appear to be
found. While the
appearance of each gyri is most probably because of gyral bias \citep{VanEssen2014},
the parcels inside them show specific patterns of structural connectivity (fig.
\ref{fig:conn_fing}), and, as seen in section 3.5.2, functional specialization.

\subsection{Our Results Show a Close Relationship Between Structural Connectivity
and Brain Function.}
We assessed the functional specialization of some of our parcels by showing
how they overlap with responses to functional and cognitive tasks measured
with fMRI. In particular, for all the studied tasks, the parcels contained
a higher proportion of positive values than negative ones as expressed by the
positive mean values reported in tables 2 and 3. For some parcels there
were not even negative values. Moreover, several of the histograms on figures
\ref{fig:function_motor} and \ref{fig:function_cognition} show a high frequency
of z-score values greater than 5, which indicate a significant correlation with
functional activation. Therefore, our results show, for some tasks, the
strong relationship between extrinsic connectivity and functional
specialization in the human brain cortex. 
\subsection{Our Parcels Are Not Similar to Those Obtained by Glasser et al. (2016) But 
			Possess Better Functional Specialization for Motor Tasks.}
Our parcels were not related to those of \citet{Glasser2016}. This is shown by the
obtained adjusted Rand index score between them (0.28). It's important to remark that
our parcels are purely based on extrinsic connectivity, meanwhile those of \citet{Glasser2016}
do not use dMRI information. Glasser's parcels are mostly based on myelin and functional information.
In particular, their subdivision of the sensori-motor cortex (green parcels in fig.
\ref{fig:glasser_and_me}) is mostly based in Myelin maps as shown in Figure 4.a of 
\citet{Glasser2016}. Because of this, their parcels in the sensori-motor cortex contain 
a wide range of z-scores when compared with responses to functional stimuli as shown by
histograms a; b and c in fig.
\ref{fig:glasser_functional}. In contrast, our parcels in the sensori-motor cortex,
for a coarser parcellation, show a good overlap with function and are in agreement with
the motor strip mapping as discussed in the previous section. Also, for the case of story
categorization; shape recognition and face recognition, our parcels show a similar
distribution of z-scores (fig. \ref{fig:function_motor}) than those with the highest 
mean z-scores of \citet{Glasser2016} (parcels d; e and f of  fig. \ref{fig:glasser_functional}).
\section{Conclusion}
Understanding how the brain is structurally organized and its relationship
with functionality is an open question in neuroscience. Recent advances in
acquisition and modeling techniques on dMRI have facilitated to study 
axonal connectivity in the brain. However, parceling the whole cortex based
on a structural criterion remained challenging. In this work we presented a 
connectivity model, framed tractography within our model and 
presented a parceling technique that allows parcellation
of the whole brain in both single subject and groupwise cases. Our
technique, along with the obtained groupwise parcellation, could have major
implications both in cognitive neuroscience and in development-aging studies.
At the same time, our technique could help to lower the gap between structural
connectivity and brain function, since some of our pure structural parcels showed
good overlapping with responses to functional tasks.

Both our parceling tool and the obtained groupwise parcellation are or will
be soon freely available in GitHub and Neurovault. These tools provide a 
sound basis for new studies on human cognition, brain development, aging 
and disease. These tools can create fine parcellations of cortical areas, 
improving our knowledge about cortical organization. Future comparison with 
functional connectivity could lead to finally unraveling the link between axonal
connectivity and brain function. \\

{\noindent \small
\textbf{Acknowledgments:} This work has received funding from the European
Research Council (ERC) under the European Union's Horizon 2020 research and
innovation program (ERC Advanced Grant agreement No 694665 : CoBCoM). This
work has received funding from the Inria Associated Team Large Brain Nets Grant.
This work also has received the following funding: NIH P41EB015902.}
\section{References}
\bibliographystyle{elsarticle-harv}
\bibliography{bibtex}
\end{document}